\documentclass[prb, reprint, showpacs, superscriptaddress]{revtex4-2}

\usepackage{graphicx} 
\usepackage{dcolumn}  
\usepackage{bm}       
\usepackage{amsmath}  
\usepackage{amssymb}  
\usepackage{hyperref} 
\usepackage{siunitx}
\hypersetup{
    colorlinks=true,
    linkcolor=blue,
    urlcolor=blue,
    citecolor=blue
}

\begin{document}

\title{Ferroelectrically Switchable Chirality in Topological Superconductivity}

\author{Kai-Zhi Bai}
\affiliation{Department of Physics, The University of Hong Kong, Pokfulam Road,
Hong Kong, China}
\author{Bo Fu}
\affiliation{School of Sciences, Great Bay University, Dongguan, China}
\author{Shun-Qing Shen}
\email{sshen@hku.hk}
\affiliation{Department of Physics, The University of Hong Kong, Pokfulam Road,
Hong Kong, China}

\date{\today}

\begin{abstract}
The interplay between ferroelectricity, magnetism, and superconductivity
provides a rich platform for discovering novel quantum phenomena.
Here, we develop an effective theory and propose a heterostructure composed of an antiferromagnetic
bilayer MnBi$_{2}$Te$_{4}$ coupled with the s-wave superconductor
Fe(Se,Te), enabling the realization of chiral topological superconductivity
(CTSC) with switchable chirality. The chirality of the CTSC is controlled
by the direction of spontaneous polarization, which arises from interlayer
sliding-induced ferroelectricity or charge transfer in the bilayer
MnBi$_{2}$Te$_{4}$. This sliding mechanism breaks the $\mathcal{M}_{z}\mathcal{T}$
and $\mathcal{PT}$ symmetries, leading to the anomalous Hall effect
in the spin-polarized metallic Dirac band and drives the emergence
of CTSC when the s-wave superconductivity appears. Our work not only
provides a new pathway to achieve and control topological superconductivity
but also opens avenues for experimental exploration of Majorana physics
and topological quantum computation.
\end{abstract}

\maketitle

\section{Introduction}

The searching for the topological superconductivity\cite{Kitaev2001Unpaired,Fu2008Superconducting,Lutchyn2010Majorana,Qi2011Topological,shen2012topological,Sato2017TSC},
particularly chiral topological superconductivity (CTSC)\cite{Read2000Paired,Wilczek2009Majorana,Qi2010Chiral,Wang2015Chiral,Fu2023Anomalous},
has been a subject of intense research interest in recent years, due
to its emergent chiral Majorana physics and potential applications
in topological quantum computation and topological quantum information
processing\cite{Bravyi2002Computation,Kitaev2003TQC,Nayak2008NonAbelian,Alicea2011TQIP,Lian2018TQC}.
However, existing approaches to realize CTSC, such as doped topological
insulators (TIs)\cite{Fu20073DTI,xia2009observation,zhang2009topological,chen2010massive}
combined with quantum anomalous Hall (QAH) systems \cite{Haldane1988QAHE,Yu2010QAHE,Chang2013QAHE},
face significant challenges. These systems require precise fine-tuning
of parameters to achieve the topologically non-trivial phase, posing
a major obstacle to experimental realization and practical applications.
Luckily, the recent discovery of the 2D ferroelectric materials\cite{Stern2021Interfacial,Tsymbal2021Ferroelectric,Wang2022rhombohedral,Weston2022FEtwisted,Gou2023ferroelectricity},
and in particular, the polar stacking bilayer MnBi$_{2}$Te$_{4}$\cite{Cao2023Switchable,Ren2022MBT,Luo2023MBT,Muzaffar2025Switchable},
offers a new opportunity to realize CTSC in a more controllable and
feasible manner. Referring to the phase with spontaneous net electric
polarization, ferroelectricity shares much similarity with ferromagnetism,
like the tunability by external fields and symmetry breaking as the
prerequisite, while differing in electric rather than magnetic, and
inversion rather than time-reversal, and leads to various functional
device applications\cite{Scott2007Ferroelectrics,Young2012Shift,Dagdeviren2014response}.
When cooperating with magnetism, multiferroic or magnetoelectric materials\cite{Eerenstein2006Multiferroic,Dong2015Multiferroic}
can be realized, with coexisting ferroic orders and even couplings
between, promising for intersecting manipulations. This brings us
to the previously mentioned multiferroic polar stacking bilayer MnBi$_{2}$Te$_{4}$,
which exhibits both ferroelectricity and antiferromagnetism, making
it a promising platform for realizing CTSC in the proximity of superconducting
pairing.

\begin{figure*}[htbp]
\centering \includegraphics[width=\textwidth]{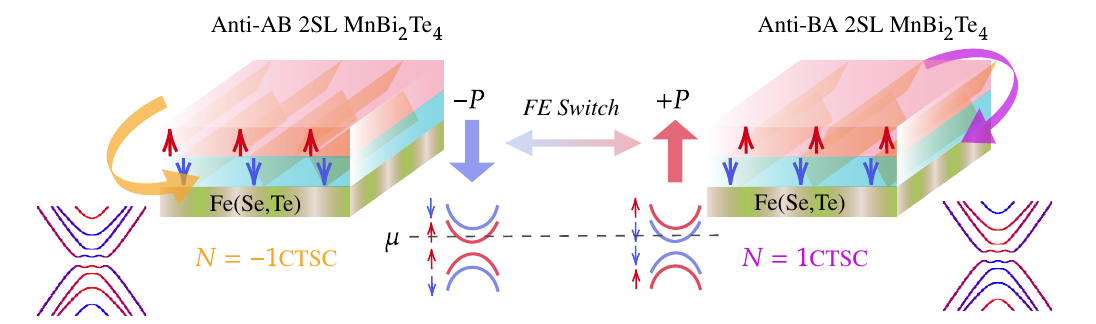} \caption{Schematic of the bilayer MnBi$_{2}$Te$_{4}$/Fe(Se,Te) heterostructure.
In the polar stacking configuration, the crystal orientation of the
bilayer MnBi$_{2}$Te$_{4}$ exhibits z-mirror symmetry, with two
stable configurations (anti-AB and anti-BA) showing ferroelectricity
due to interlayer sliding. Both configurations maintain interlayer
antiferromagnetism and exhibit spin-splitting in the band structure.
When coupled with Fe(Se,Te), chiral topological superconductivity
(CTSC) emerges as the chemical potential lies within a single spin
band. The chirality of the CTSC can be switched by the direction of
ferroelectric polarization.}
\label{fig:schematic}
\end{figure*}

In this work, we propose the heterostructure of a polar stacking antiferromagnetic
bilayer MnBi$_{2}$Te$_{4}$ in close proximity to an $s$-wave superconductor
Fe(Se,Te) \cite{He2014interface,Owada2019Electronic,Qin2020Superconductivity,Ding2022Observation,Yuan2024Coexistence}.
As shown in Fig.~\ref{fig:schematic}, when the Fermi surface intersects
a single band to form an isolated Fermi loop, this system exhibits
CTSC with switchable chirality, controlled by the spontaneous polarization
direction arising from the ferroelectricity induced by interlayer
spin-orbital coupling due to sliding in the magnet. As we will demonstrate,
this prospect is made possible by the deep cooperation between inversion
$\mathcal{P}$-breaking ferroelectricity and antiferromagnetism in
the system. We further endow a guideline for realizing of CTSC in
superconducting multiferroic materials, where the Chern number $N$
of the superconducting states is determined by the residual chirality
summation over all the Fermi loops: $N=\sum_{i}|n_i|\mathrm{sgn}(\sigma_{H}^{i})$.
The residual chirality of Fermi loop for band $i$, denoted by $|n_i|\mathrm{sgn}(\sigma_{H}^{i})$,
encodes the Fermi loop winding number with its sign of the anomalous Hall conductivity arising from all
occupied states of this band, thus dictating the magnitude and direction of the
Hall current\cite{haldane2004berry,Xiao2010Berry,Nagaosa2010AHE}.

In the following, we start by introducing the polar-stacking structure and microscopic mechanism for ferroelectricity in bilayer MnBi$_{2}$Te$_{4}$ in Section \ref{sec:ferroelectricity},
then we demonstrate the switchable anomalous Hall effect (AHE) in the normal state in Section \ref{sec:AHE}.
After that, we introduce superconductivity in the heterostructure and demonstrate the emergence of CTSC in Section \ref{sec:CTSC}.
We discuss the coexistence of AHE and CTSC in Section \ref{sec:coexistence} and illustrate the underlying mechanisms in Section \ref{sec:singleloop} and \ref{sec:deformation}.
Finally, we propose to detect possible CTSC phase with temperature-dependent thermal Hall conductivity in Section \ref{sec:THC}.

\section{Effective theory for ferroelectricity in polar-stacking bilayer $\textbf{MnBi}_{\mathbf{2}}\textbf{Te}_{\mathbf{4}}$}\label{sec:ferroelectricity}

\begin{figure}[htbp]
\centering \includegraphics[width=8.6cm]{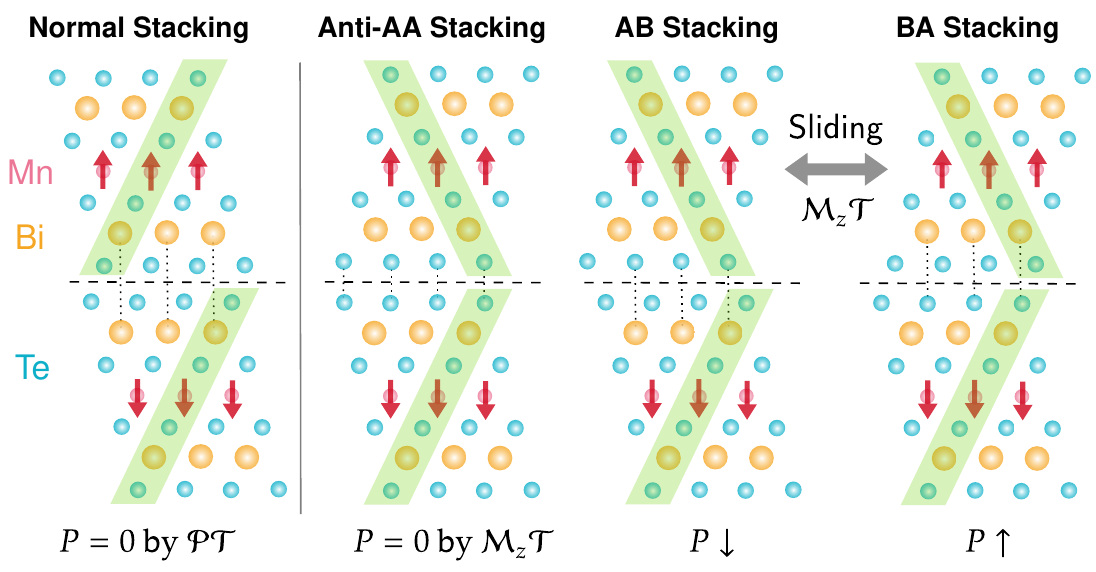} \caption{Comparison of atomic structures between the normal and the stable
polar stacking of bilayer MnBi$_{2}$Te$_{4}$ (side views). We utilize
green stripe to denote the crystal orientation in each SL. From left
to right, the normal stacking, anti-AA, AB and BA stacking are depicted.}
\label{fig:comparestack}
\end{figure}

The polar-stacking structure of bilayer MnBi$_{2}$Te$_{4}$ arises
from a mirror-twin boundary\cite{Cao2023Switchable}, where the bottom
septuple layer acts as a mirror image of the top layer across the
$z=0$ plane, accompanied with a further time-reversal operation that
reverses the magnetic order and leads to the interlayer antiferromagnetism.
This anti-AA stacking phase with its atomic structure is presented
in Fig.~\ref{fig:comparestack}. First-principle calculations\cite{Cao2023Switchable}
reveal four bands near the $\Gamma$ point, primarily composed of
the $p$-orbitals of Bi and Te atoms and are well-separated from others,
making it feasible to construct an effective $\bm{k}\cdot\bm{p}$
model for the system. In that sense, we consider a $p_{z}$ orbital-mixed
basis $|P_{z},\uparrow/\downarrow\rangle$ for the four bands, with
$\uparrow/\downarrow$ representing the spin-up/down states. Using
the theory of invariants \cite{winkler2003spin,zhang2009topological,Liu2010Model,Yang2014Classification,Acosta2018Tightbinding,fu2024invariants},
the low-energy effective Hamiltonian is given by $H_{AA}=H_{0}+H_{\text{\text{AFM}}}$,
with
\begin{equation}
H_{0}(\bm{k})=\varepsilon(k)+\lambda\tau_{z}(\bm{k}\times\bm{\sigma})_{z}+m(k)\tau_{x}+\tau_{y}\bm{d}(\bm{k})\cdot\bm{\sigma},
\end{equation}
where $\varepsilon(k)=\varepsilon_{0}+bk^{2}$, $m(k)=m_{0}+tk^{2}$,
$(d_{x},d_{y})(\bm{k})=\omega(k_{x}^{2}-k_{y}^{2},-2k_{x}k_{y})$,
respectively. Here, $\tau$ and $\sigma$ are the Pauli matrices acting
on the layer and spin spaces, with $\lambda$ and $m$ reading as
the spin-orbital and finite-thickness coupling strengths. The last
term, which we call as the quadratic warping term \cite{Fu2009Hexagonal},
is allowed as the second harmonic of three-fold $z$-rotation $C_{3z}$
symmetry, attributing to the absence of the inversion symmetry $\mathcal{P}$,
which is replaced by a $z$-mirror symmetry $\mathcal{M}_{z}$. Other
symmetries for the construction include the time-reversal $\mathcal{T}$
and the $x$-mirror $\mathcal{M}_{x}$ symmetries. Due to the ferromagnetic
intra-layer and antiferromagnetic inter-layer couplings, the second
part 
\begin{equation}
    H_{\text{AFM}}=g\tau_{z}\sigma_{z}
\end{equation}
is introduced for magnetization,
which preserves both $\mathcal{M}_{z}\mathcal{T}$ and $\mathcal{PT}$
symmetries.

\begin{figure}[htbp]
\centering \includegraphics[width=8.6cm]{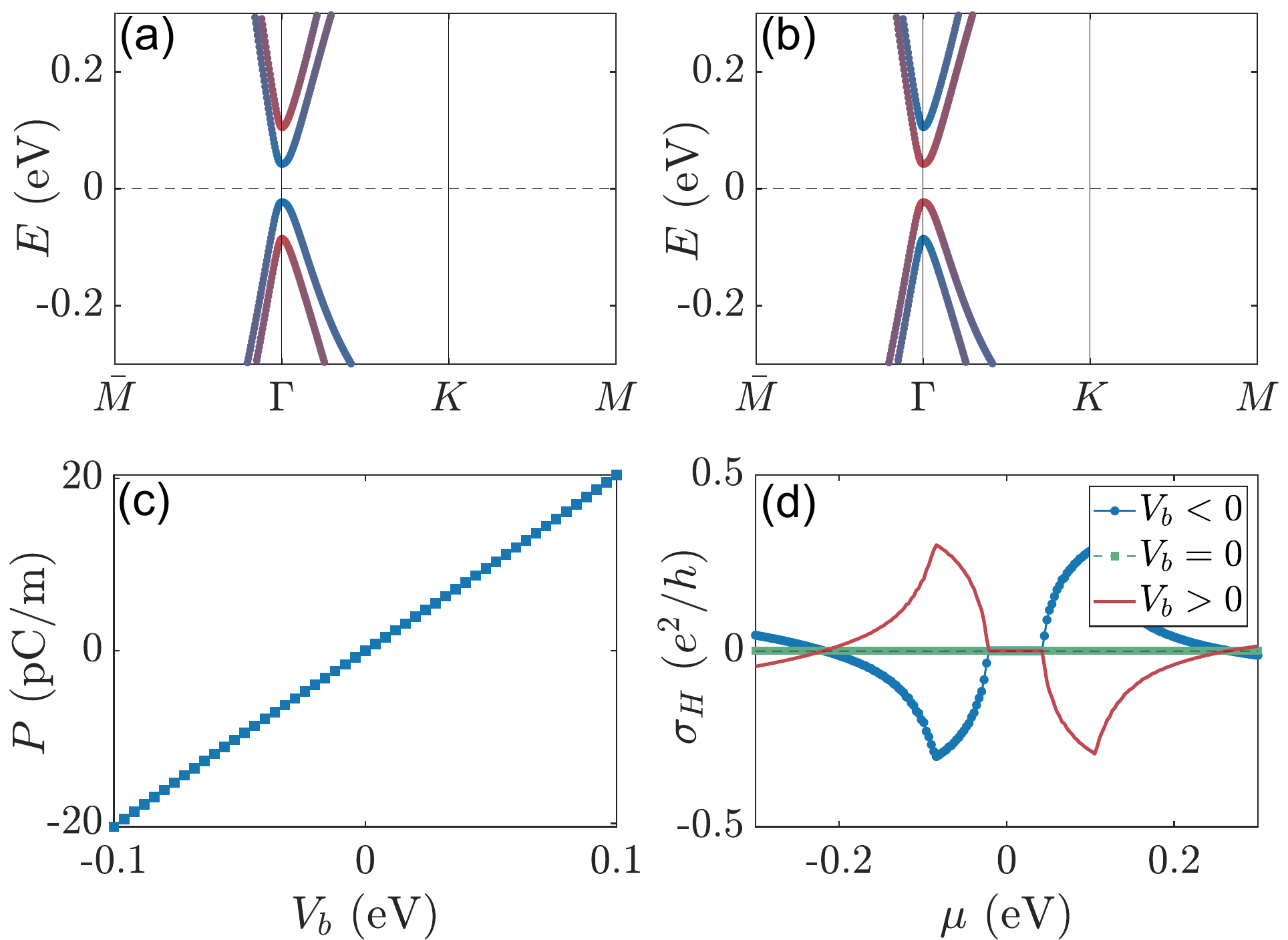} \caption{Band structure, spontaneous polarization and anomalous Hall effect
in polar stacking bilayer MnBi$_{2}$Te$_{4}$. (a) (b) Band structures
of the BA and AB stacking phases, with the color representing the
spin polarization. The electric polarizations in the two phases are
positive and negative, respectively. (c) Spontaneous polarization
$P_{z}$ as a function of the ferroelectricity potential $V_{b}$.
(d) Anomalous Hall conductance as a function of the chemical potential
of the AB and BA stacking phases, with anti-AA stacking phase serving
as a reference.}
\label{fig:band_hall_pol}
\end{figure}

Finally, we let the system relax to its thermodynamically stable anti-AB
or BA stacking phase, where the inter-layer sliding occurs shown schematically
in Fig.~\ref{fig:schematic}. The realistic atomic structures are
depicted in Fig.~\ref{fig:comparestack}. This introduces the ferroelectric
term 
\begin{equation}
H_{\text{FE}}=V_{b}\tau_{z}\sigma_{0},
\end{equation}
which breaks $\mathcal{M}_{z}\mathcal{T}$ and $\mathcal{PT}$ simultaneously.
Theoretically, this signature is evident when considering the perturbative
effect of $V_{b}$ over the polarization at $\Gamma$ point. Since
$H(k=0)=\varepsilon_{0}+m_{0}\tau_{x}+g\tau_{z}\sigma_{z}+V_{b}\tau_{z}$,
its eigenstates $|\pm,s\rangle$ with $s$ eigenvalue of $\sigma_{z}$
lead to polarization $P_{z}(0)\propto-e\sum_{s}\langle\tau_{z}\rangle_{-,s}=-e\sum_{s}\cos\theta_{s}$
aside from a length constant, where $\cos\theta_{s}=\frac{sg+V_{b}}{\sqrt{m_{0}^{2}+(sg+V_{b})^{2}}}$.
For illustration, considering a weak $V_{b}$, we have $P_{z}(0)\propto\frac{2eV_{b}}{\tilde{m}_{0}}$
with $\tilde{m_{0}}=m_{0}(1+(\frac{g}{m_{0}}))^{3/2}$, which is directly
proportional, thus sign alignment and tunability. To perform numerical
calculations, we put this model on a hexagonal lattice as 
\begin{equation}
H^{h}(\bm{k})=H_{0}^{h}(\bm{k})+H_{\text{AFM}}+H_{\text{FE}},
\end{equation}
with the form of $H_{0}^{h}$ presented in Eq.~(S10) of Ref.\cite{Note-on-SM}.
As presented in Fig.~\ref{fig:band_hall_pol}(c), the total polarization
$P_{z}$ is almost a linear function of $V_{b}$ in a large regime,
and confirms the theoretical prediction. Notably, when $V_{b}=0$,
we have $P=0$ as for a non-polarized system, representing a ferroelectric
phase transition point. Note that $V_{b}$ is not a monotonic, but
a periodic function of the interlayer sliding distance $\delta$,
and for an oversimplified illustration, we can propose $V_{b}(\delta)=V_{b}^{0}\sin(2\pi\delta/a)$
thus $P=P_{0}\sin(2\pi\delta/a)$ in the linear regime.

As detailed in Sec. SII of Ref.\cite{Note-on-SM}, we construct a
thin-film model of the bilayer system and project it onto the four
lowest energy bands near the band edges \cite{Bai2023Metallic,Bai2024Dirac}.
This reveals that the ferroelectricity term arises from interlayer
spin-orbital coupling (SOC), induced by a non-trivial sliding mechanism.
In the effective model, this term is referred to as the ferroelectric
potential (FEP) term, distinct from external electric fields or gate-voltage
effects. Physically, this additional SOC term interacts with the original
$z$-SOC and Dirac mass of the bilayer system in a non-commutative
manner, creating an imbalance in the dynamics between the two layers
for both spin components. This imbalance leads to charge transfer
and the emergence of ferroelectricity.

\section{Switchable anomalous Hall effect}\label{sec:AHE}

The interplay between antiferromagnetism and ferroelectricity in the
bilayer MnBi$_{2}$Te$_{4}$ leads to a non-vanishing anomalous Hall
effect (AHE). While interlayer sliding does not alter the band topology,
the system exhibits zero Hall conductance when the chemical potential
lies within the gap \cite{Cao2023Switchable,Ren2022MBT,Luo2023MBT}.
However, the ferroelectricity term $V_{b}$ breaks the $\mathcal{PT}$
symmetry, enabling the system to display an AHE when the chemical
potential is shifted into the valence or conduction bands. This is
confirmed by our explicit calculations of the Hall conductivity using
the Kubo formula in terms of the Berry curvatures \cite{mahan2000many}.
As shown in Fig. 3(d), the sign of the Hall conductance changes with
different stacking configurations at a fixed Fermi level. To understand
this behavior, we examine the $\Gamma$ point of the four band edges,
given by
\begin{equation}
    E_{s,\pm}(0)=\varepsilon_{0}\pm\sqrt{m_{0}^{2}+(sg+V_{b})^{2}},
\end{equation}
where $s$ denotes the spin index of $\sigma_{z}$. In the valence
bands, the spin-up band has a higher energy when $V_{b}<0$ (with
$g>0$ by default), corresponding to the anti-AB stacking phase, and
vice versa. This results in the band alignment sequences $\downarrow$-$\uparrow$-$\uparrow$-$\downarrow$
for the anti-AB stacking and $\uparrow$-$\downarrow$-$\downarrow$-$\uparrow$
for the anti-BA stacking, as shown in Fig. 3(a)(b). The spin-dependent
band splitting leads to a non-compensated Berry curvature integral,
resulting in the observed AHE in the conduction and valence bands.
For instance, when the chemical potential lies within the higher valence
band, the fully occupied lower energy bands contributes no Hall effect,
while the higher energy band contributes a finite Hall conductance
due to the Berry curvature integral \cite{haldane2004berry,Bai2024Dirac},
with its sign determined by the spin polarization. Notably, when $V_{b}=0$,
corresponding to the non-polarized anti-AA stacking phase, the valence
and conduction bands become degenerate, and consequently the Hall
conductance vanishes. This attachment and separation of valence bands
serve as a signature of the ferroelectric phase transition in this
magnetic system, providing a potential experimental marker for ferroelectricity.

\section{Chiral topological superconductivity in $\textbf{MnBi}_{\mathbf{2}}\textbf{Te}_{\mathbf{4}}\textbf{/Fe(Se,Te)}$ heterostructure}\label{sec:CTSC}

\begin{figure}[htbp]
\centering \includegraphics[width=8.6cm]{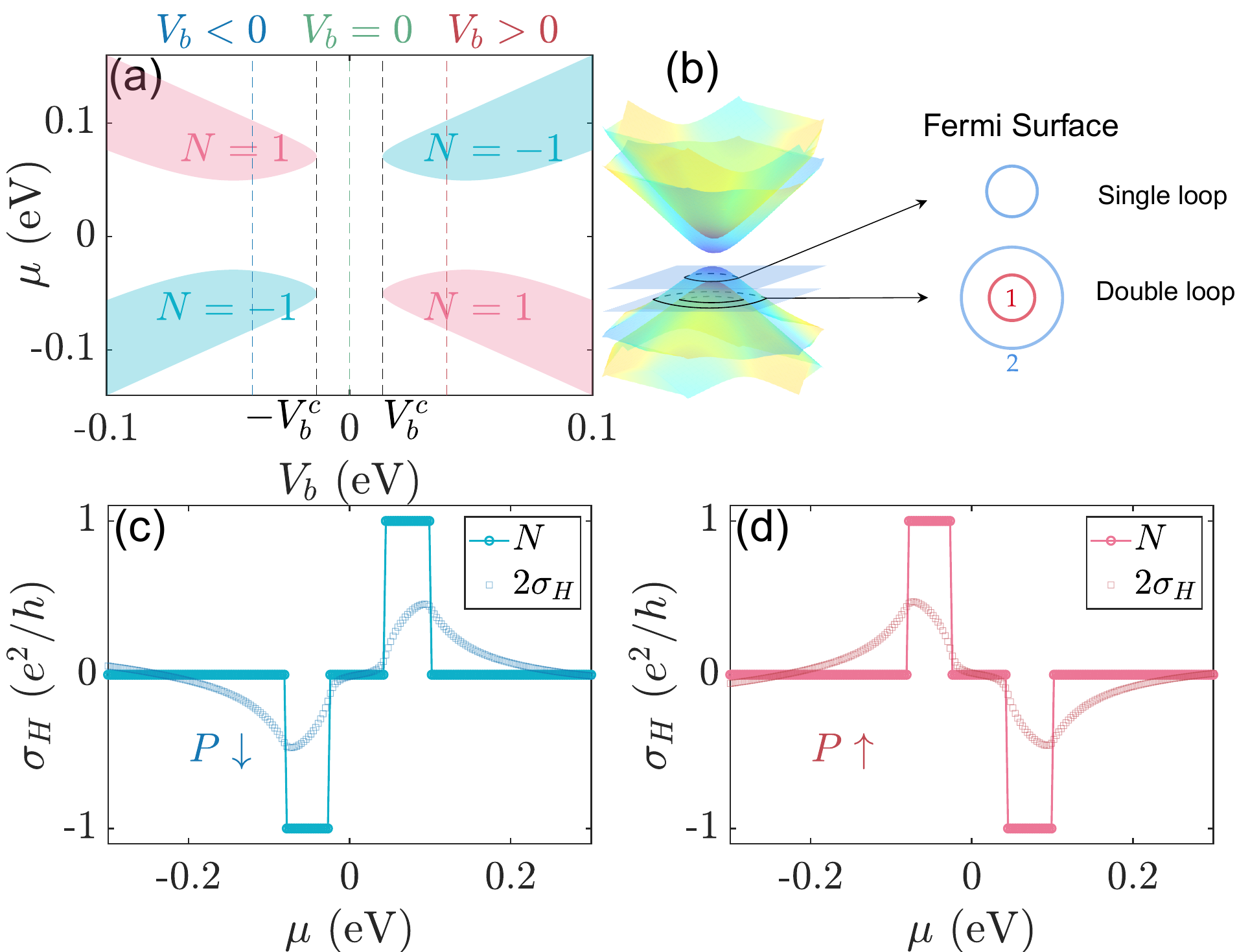} \caption{Topological phase diagram (a), a schematic diagram illustrating the
how the position of chemical potential influence the Fermi surface
loop structure (b) and superconducting Chern number $N$ and anomalous
Hall conductance $\sigma_{H}$ for (c) AB and (d) BA stacking cases
in bilayer MnBi$_{2}$Te$_{4}$/Fe(Se,Te) heterostructure. In the
phase diagram of (a), the three $V_{b}$-constant lines indicate the
three stacking manners. We also define $V_{b}^{c}=|\Delta m_{0}|/\sqrt{g^{2}-\Delta^{2}}$
as the critical FEP intensity for the realization of the topological
superconductor.}
\label{fig:phase_tsc}
\end{figure}

We now consider the heterostructure of bilayer MnBi$_{2}$Te$_{4}$
on a few-layer Fe(Se,Te), where superconductivity has been observed
at the interface closely related experiment \cite{Yuan2024Coexistence}.
Proximity to the s-wave superconductivity in Fe(Se,Te) \cite{He2014interface,Owada2019Electronic,Qin2020Superconductivity,Ding2022Observation,Yuan2024Coexistence}
induces a pairing potential in the bilayer MnBi$_{2}$Te$_{4}$, described
by the Bogoliubov-de Gennes (BdG) Hamiltonian: 
\begin{equation}
H_{\text{BdG}}(\bm{k})=\rho_{z}\left(H_{0}(\bm{k})+H_{\text{FE}}-\mu\right)+\rho_{x}\Delta\tau_{0}+\rho_{0}H_{\text{AFM}},
\end{equation}
under the Nambu spinor basis, where $\mu$ is the chemical potential
and the Pauli matrices $\rho$ represent the particle-hole space.
For simplicity we assume a constant $s$-wave pairing potential $\Delta$
in the bilayer. Due to the $\mathcal{T}$ and $\mathcal{P}\mathcal{T}$
symmetry breaking, our particle-hole symmetric ($\mathcal{C}=\rho_{y}\sigma_{y}\mathcal{K}$)
$D$-class system is allowed to carry a non-trivial topological index,
namely the superconducting Chern number $N$\cite{Kitaev2009Periodic,Ryu2010tenfold,Chiu2016Classification}.
The possibility for the emergence of CTSC in the system is reflected
by the phase boundaries defined by $\det(H_{\text{BdG}}(0))=0$, which
leads to the equation that
\begin{equation}
    \sqrt{\tilde{\mu}^{2}+\Delta^{2}}=f_{\pm}(\Delta),
\end{equation}
where the shifted chemical potential $\tilde{\mu}=\mu-\varepsilon_{0}$,
$f_{\pm}(\Delta)=\sqrt{\tilde{m}_{0}^{2}+g^{2}\pm2\sqrt{g^{2}V_{b}^{2}-\Delta^{2}\tilde{m}_{0}^{2}}}$
with $\tilde{m}_{0}^{2}=m_{0}^{2}+V_{b}^{2}$. For the expression
to make sense, we require $g^{2}V_{b}^{2}>\Delta^{2}\tilde{m}_{0}^{2}$
and $\tilde{m}_{0}^{2}+g^{2}>2\sqrt{g^{2}V_{b}^{2}-\Delta^{2}\tilde{m}_{0}^{2}}$,
which can be usually satisfied due to the smallness of $\Delta$.
Under the situation, the CTSC phase appears when the self-consistent
inequality $f_{-}(\Delta)<\sqrt{\tilde{\mu}^{2}+\Delta^{2}}<f_{+}(\Delta)$
is satisfied, which is shown in the phase diagram in Fig.~\ref{fig:phase_tsc}(a),
where two chemical potential windows emerge for each FEP intensity
satisfying the above requirement. Especially, for $\Delta\rightarrow0$,
we have the windows as $\sqrt{m_{0}^{2}+(|g|-|V_{b}|)^{2}}<|\tilde{\mu}|<\sqrt{m_{0}^{2}+(|g|+|V_{b}|)^{2}}$,
which cannot exist either without the ferroelectricity or antiferromagnetism.
In this limit, an arbitrarily small $V_{b}$ can lead to the realization
of a single Fermi loop, leading to the CTSC phase. Out of the window,
the double Fermi loops appear and CTSC disappears. Turning to the
practical phases, we see that for the anti-AA stacking phase, $V_{b}=0$
and the Fermi surface always consists of two Fermi loops, consequently
only $N=0$ trivial phase is observed when the superconducting gap
open. While on the other hand, for the anti-AB/BA stacking phases
shown in Fig.~\ref{fig:phase_tsc}(c)(d), the superconducting Chern
number $N$ changes its sign with the ferroelectric potential $V_{b}$
at a fixed Fermi level inside the observable windows, leading to a
ferroelectrically switchable $N=\pm1$ CTSC phase.

\section{Coexistence of anomalous Hall effect and topological superconductivity}\label{sec:coexistence}

The coexistence of the anomalous Hall effect and topological superconductivity
in the system is particularly intriguing. Notably, the anomalous Hall
conductance remains non-zero and non-quantized even when the system
is gapped by superconductivity \cite{Takahashi2002Hall,Sacramento2012Anomalous,POjanen2013Anomalous,Sacramento2014Hall,Bednik2016Anomalous}.
The Chern number $N$ is determined by the Berry curvature integrals
of the occupied bands, which can be re-written using the Kubo formula
evaluated over the quasi-particle current $j_{\text{QP}}^{\mu}=-e\partial H_{\text{BdG}}/\hbar\partial k_{\mu}$.
Meanwhile, the anomalous Hall conductivity $\sigma_{H}$ reflects
the retarded electronic current response to an external electric field,
which is determined by the electric correlation functions and also
manifests itself as the Kubo formula, which is however evaluated over
the electronic charge current $j_{\text{el}}^{\mu}=-\delta H_{\text{BdG}}/\delta A_{\mu}$,
and is modified to contain only half the magnitude to compensate the
Nambu space doubling. Since these two phenomena represent different
physical quantities, their coexistence is not contradictory. Moreover,
the superconducting pairing is typically weak, so its effect on the
AHE (which does not relate to a specific topological invariant in
this case) is perturbative. This allows both AHE and TSC to coexist
without mutual interference, providing a unique platform for studying
the interplay between these phenomena

Written under the BdG form, one can show rather generally that for
the $s$-wave pairing 
\begin{equation}
    j_{\text{el}}^{\mu}=\rho_{z}j_{\text{QP}}^{\mu},
\end{equation}
indicating their relations and differences. Specifically, when we
separate $j_{\text{QP}}=j_{e}+j_{h}$ into the electron and hole parts
(a direct sum indeed), at the same time we have $j_{\text{el}}=j_{e}-j_{h}$,
which enlightens us to write
\begin{equation}
    \sigma_{H}=(\sigma_{H}^{ee}+\sigma_{H}^{hh}-\sigma_{H}^{eh}-\sigma_{H}^{he})/2,
\end{equation}
with each part evaluated using unmodified Kubo formula over the corresponding
currents. Under the limit $\Delta\rightarrow0$ when the electron-hole
mixing fades away, leading to $\sigma_{H}=\sigma_{H}^{ee}$. Notice that the
superconducting Chern number
\begin{equation}
    N=\sigma_{H}^{ee}+\sigma_{H}^{hh}+\sigma_{H}^{eh}+\sigma_{H}^{he}
\end{equation}
contains a singular limit, i.e., $N(\Delta\rightarrow0)\in\mathbb{Z}$
is an integer, while $N(\Delta=0)=2\sigma_{H}^{ee}$ gives twice the
anomalous Hall conductance $\sigma_{H}$ of the normal state.

For a more intuitive picture of this coexistence, we may consider
the net electron occupation number in the Bardeen-Cooper-Schrieffer
(BCS) ground state $|\text{BCS}\rangle=\prod_{k}(u_{k}+v_{k}c_{k\uparrow}^{\dagger}c_{-k\downarrow}^{\dagger})|0\rangle$
relative to the electronic insulator background $|\text{EI}\rangle=\prod_{k}c_{k\uparrow}^{\dagger}c_{-k\downarrow}^{\dagger}|0\rangle$,
with the same $\prod_{k}$ structure and energy spectrum, which reads
$N_{e}=\sum_{k,\sigma=\uparrow\downarrow}\langle c_{k\sigma}^{\dagger}c_{k\sigma}\rangle_{\text{BCS}}-\langle c_{k\sigma}^{\dagger}c_{k\sigma}\rangle_{\text{EI}}=-\sum_{k}|u_{k}|^{2}$
and departures from zero. This implies that when the system is insulating
with the chemical potential touching one band edge, the BCS state
is indeed a non-fully occupied state for the electrons, allowing for
the non-vanishing AHE. A more detailed discussion is presented in
Sec. SV of \cite{Note-on-SM}.

\section{Single loop anomalous Hall effect leads to chiral topological superconductivity}\label{sec:singleloop}

Single or double Fermi loop of the non-superconducting system ($\Delta=0$)
determines topological property of the induced superconductivity ($\Delta\neq0$).
The discussion below focuses on the limit of $\Delta\rightarrow0$,
but is not limited to that. The key lies in a first idealization and
then a following continuous deformation of the band structure, and
the fact that the topological invariant remains unchanged if the band
gap does not close and reopen. When the Fermi surface crosses a single
band, creating a single Fermi loop as presented in right-top of Fig.~\ref{fig:phase_tsc}(b),
one idealizes this crossed band (and its partner if necessary) into
a regulated massless Dirac cone resembling a single surface Dirac
cone of semi-magnetic TI film, which is known to carry a half-quantized
anomalous Hall conductance $\sigma_{H}^{\text{MD}}=\pm e^{2}/2h$\cite{fu2022half,Zou2023half,Bai2024Dirac,fu2024invariants,Shen2024Coshare}
originated from the $\pi$ Berry phase around its Fermi loop by Stokes
theorem. Importantly, this half-quantized Hall conductivity, evolves
into a quantized superconducting Chern number $N=\mathrm{sgn}(\sigma_{H}^{\text{MD}})$
when the system is gapped by superconductivity, $\Delta\neq0$ \cite{Fu2023Anomalous},
i.e., $N=\mathrm{sgn}(\sigma_{H}^{\text{MD}})=2\sigma_{H}^{\text{MD}}$
where the latter equality holds only in the limit of $\Delta\rightarrow0$.
We then let the system evolve continuously from the massless Dirac
cone to the massive Dirac cone, and since the superconducting band
gap at the Fermi surface survives the deformation, the quantized Chern
number $N$ remains unchanged, $N=\mathrm{sgn}(\sigma_{H})\neq2\sigma_{H},$
while the second equality does not holds anymore due to the derivation
from ideal massless to massive Dirac cone, $\sigma_{H}\neq\pm e^{2}/2h$.
The first equality reveals the identification between superconducting
Chern number and the residual chirality. Thus in general, for a single
Fermi loop, the superconducting Chern number is $N=\mathrm{sgn}(\sigma_{H})$.
On the other hand, when the Fermi level crosses two bands, creating
a double Fermi loop labelled by $1$, $2$ as presented in right-bottom
of Fig.~\ref{fig:phase_tsc}(b), the idealization and deformation
procedure tells that $N=\mathrm{sgn}(\sigma_{H}^{1})+\mathrm{sgn}(\sigma_{H}^{2})=0,$
since the two bands carry opposite AHE, leading to the residual chirality
cancellation in the two bands. See next section for a detailed and generalized 
explanation and 
\cite{Note-on-SM} SIII.4 for analytic $p$-wave superconductivity illustration.

\section{Continuous deformation from massless HQHE to gapped AHE}\label{sec:deformation}

Going beyond the specific system, here we provide a detailed explanation of the continuous deformation from massless Dirac cone to gapped AHE.
In this section, we modulate anomalous Hall conductivity (AHC) with von Klitzing constant $e^2/h$.

\begin{figure*}[htbp]
    \centering \includegraphics[width = \textwidth]{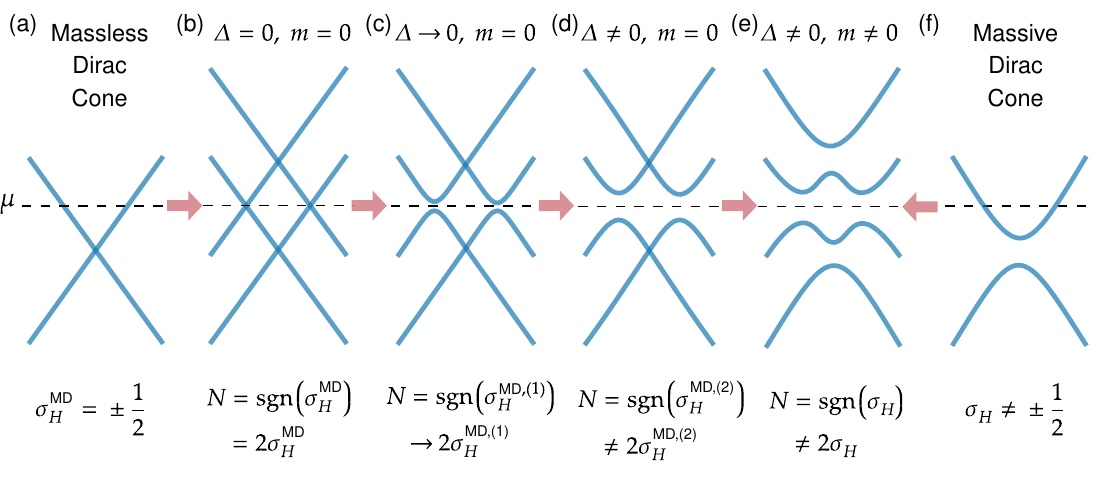} 
    \caption{CTSC evolving from that in HQHE to that in gapped AHE.
        From the left to right, we show (a) normal state for a regulated massless Dirac cone, (b) particle-hole doubled states for regulated massless Dirac cone,
        (c) (d) superconducting state for regulated massless Dirac cone (middle two), (e) superconducting state for regulated massive Dirac cone, and (f) normal state for a regulated massive Dirac cone.
        }
    \label{fig:AHECTSC}
\end{figure*}

For single Fermi loop, by allowing superconducting gap opened at the Fermi surface,
and to perform the idealization, we first extract the relevant Dirac cone out of the rest of the bands without any coupling to the other bands, 
during which we further require no band gap closing and reopening.
Then we can continuously deform the Dirac cone into a regulated massless/gapless Dirac cone, described by the Hamiltonian\cite{Bai2024Dirac}
\begin{equation}
    H_{\text{BdG}}^{\text{MD}}(\bm{k}) = (\lambda \bm{k}\cdot \bm{\sigma} - \mu) \rho_z + m(k)\sigma_z + \Delta(k)\rho_x,
\end{equation}
where $\lambda$ is the Dirac velocity, the mass term $m(k) = \Theta(-\mathrm{sgn}(b)m_0(k))m_0(k)$, $m_0(k) = m_0 - bk^2$ ($m_0b > 0$), and the pairing 
term is peaked at the Fermi surface.

The workflow is shown in Fig.~\ref{fig:AHECTSC}.
Start from leftmost (a), we require $\mu$ lies within the linear part, making the 
normal system $H^{\text{MD}} = \lambda \bm{k}\cdot \bm{\sigma} + m(k)\sigma_z$ carrying a half quantum Hall effect (HQHE)
\begin{equation}
    \sigma^{\text{MD}}_H = \frac{1}{2} \mathrm{sgn}(b),
\end{equation}
where the sign depends on the chirality $\mathrm{sgn}(b)$ carried by the massless Dirac cone consistent with the original massive Dirac cone.

Moving one step right to (b), the Nambu space is introduced without superconducting gap, and $N$ (here not Chern number) 
follows as 
\begin{equation}
    N = \mathrm{sgn}(\sigma^{\text{MD}}_H) = 2\sigma^{\text{MD}}_H = \mathrm{sgn}(b).
\end{equation}
This stands since $N = \sigma^{\text{MD},ee}_H + \sigma^{\text{MD},hh}_H + \sigma^{\text{MD},eh}_H + \sigma^{\text{MD},he}_H$
is reduced by $\sigma^{\text{MD},eh/he}_H(\Delta = 0) = 0$, since there is no particle-hole mixing, and $\sigma^{\text{MD},ee/hh}_H (\Delta = 0) = \sigma^{\text{MD}}_H$.

One step further to (c), an infinitesimal superconducting gap $\Delta$ is opened up near the Fermi surface.
We can then cut the band structure into two parts, one low-energy part as a round centered at $k = 0$ including the Fermi surface 
(and its neighboring states with non-zero pairing), and the other high-energy part correspondingly.
Under the situation, the high-energy part of the band is indeed un-affected by the pairing, with particle-hole decoupled wavefunction $(u,0)^T$ for particle and $(0,v)^T$ for hole still,
and time-reversal breaking non-zero mass term only appears in this part, which still leads to $\mathrm{sgn}(b)$ contribution to $N$ and $\sigma_H$ as in normal band structure.
The time-reversal unbroken low-energy part will develop a mixed $(u,v)^T$ wavefunction for occupied states, which, however, can be further decoupled into direct product
form of 
\begin{equation}
    \phi_{s}(\bm{k}) = \frac{1}{\sqrt{2}} \left( \begin{array}{c}
        1 \\ s e^{i\varphi_k}
    \end{array} \right)\otimes \left(\begin{array}{c}
        \cos(\theta_{s,k}/2) \\ \sin(\theta_{s,k}/2)
    \end{array}\right),
\end{equation}
with $s = \pm$, $\varphi_k$ the azimuthal angle of $\bm{k}$, and $\theta_{s,k}$ the polar angle in particle-hole space by $\cos \theta_{s,k} = (s\lambda k - \mu)/E_s(k)$, $E_s(k) = \sqrt{(s\lambda k - \mu)^2 + \Delta^2(k)}$.
This winding--particle-hole decoupled form will not contribute to either $N$ or $\sigma_H$.
However, since the band structure is gapped, $N$ now shares the meaning of superconducting Chern number, and we write
\begin{equation}
    N = \mathrm{sgn}(\sigma^{\text{MD},(1)}_H) \leftarrow 2\sigma^{\text{MD},(1)}_H,
\end{equation}
with upper $(1)$ index indicating the change.

Another step right to (d), we now relax the pairing distribution to deform into its possible configuration (for instance, a uniform constant),
with the requirement that no band gap closing and reopening near Fermi level is allowed.
Due to the topological protection by the superconducting gap, the Chern number remains unchanged, while the AHC 
is generally not half-quantized, since it is not a topological invariant in the superconducting state, and we write
\begin{equation}
    N = \mathrm{sgn}(\sigma^{\text{MD},(2)}_H) \neq 2\sigma^{\text{MD},(2)}_H,
\end{equation}
with upper $(2)$ index indicating a second change of massless Dirac cone.

For the second last step (e), we let the Dirac cone acquire its low-energy mass to be a massive Dirac cone, without band gap closing and reopening near Fermi level.
Similarly, 
\begin{equation}
    N = \mathrm{sgn}(\sigma_H) \neq 2\sigma_H,
\end{equation}
representing our conclusion for a superconductively gapped massive Dirac cone.
Finally, we embed the relative bands back to the original band structure, where the superconducting Chern number remains unchanged.

\begin{figure}[htbp]
    \centering \includegraphics[width = 8.6cm]{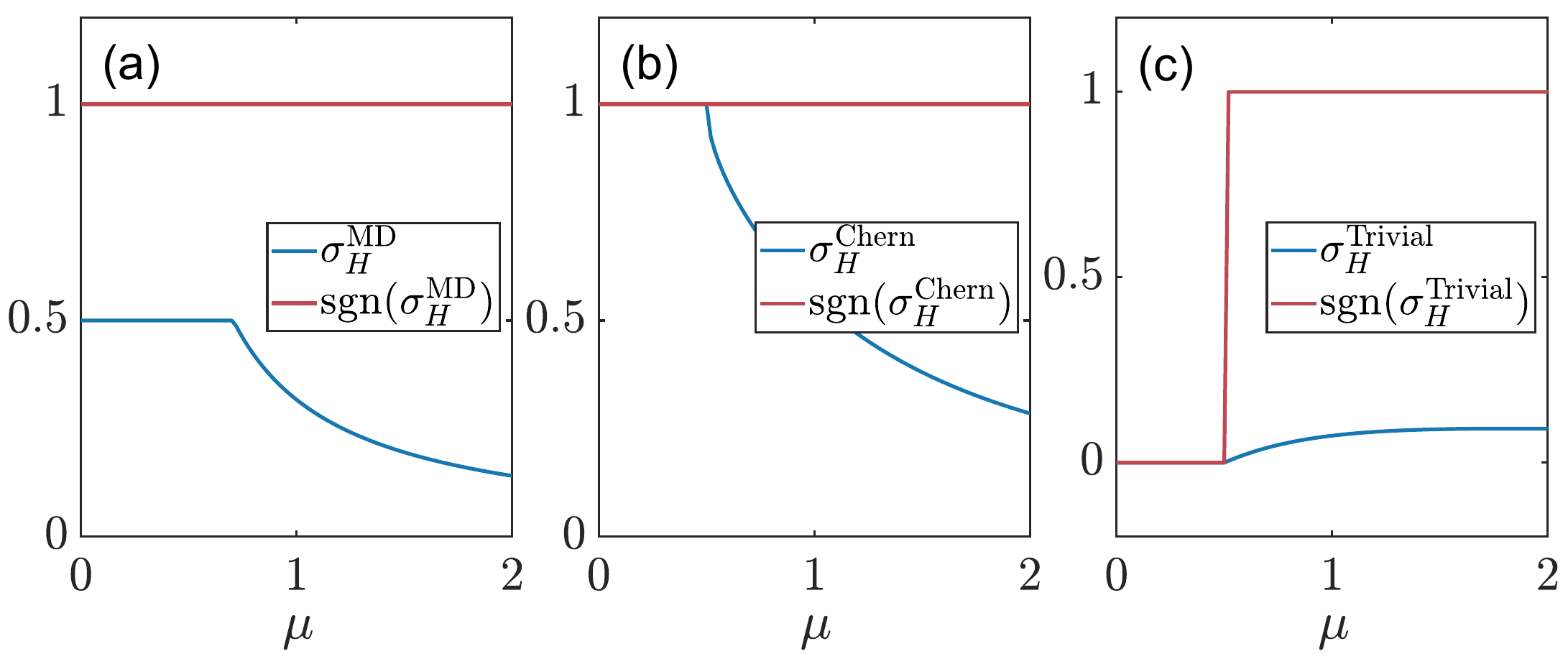} 
    \caption{AHC and its sign for (a) massless Dirac cone, (b) non-trivial massive Dirac cone, and (c) trivial massive Dirac cone, varying with the chemical potential.
    Specifically, we choose parameters $\lambda = 1$, $b = 1$, $m_0 = 0.5$, and mass term $m(k)$ for (a), $m_0(k)$ for (b) and $-m_0 - bk^2$ for (c).
    }
    \label{fig:Hall_sign_MD_Chern_Trivial}
\end{figure}

The obtained result is robust with the variation of the chemical potential, as shown in Fig.~\ref{fig:Hall_sign_MD_Chern_Trivial},
where we select massless, non-trivial massive, and trivial massive Dirac cones respectively, for illustration.
When $\mu$ lies with a single Fermi loop, the sign of AHC is uniquely determined by the residual chirality of the Dirac cone fixed by $\mathrm{sgn}(b)$,
without changing with $\mu$.
Note for the massive case, when $\mu$ is in mass gap with no loop, the sign has no meaning as superconductivity will not enter the band.
The case when $\Delta$ exceeds $m_0$ should be understood to evolve from an infinitesimal pairing at a single loop.

For the case of double Fermi loop, we employ the same strategy to extract and decouple
the relative bands into two decoupled Dirac cones.
Although this may superficially promote the symmetry of the system,
as we focus on the superconducting Chern number, we only ask for the final summation 
of Chern numbers from the superconductively gapped Dirac cones, and the value of each is not important.
Then similarly, for the idealized paired massless Dirac cones, we have 
\begin{equation}
    \begin{split}
        N_1 &= \mathrm{sgn}(\sigma^{\text{MD},1}_H) = 2\sigma^{\text{MD},1}_H = 1,\\
        N_2 &= \mathrm{sgn}(\sigma^{\text{MD},2}_H) = 2\sigma^{\text{MD},2}_H = -1,
    \end{split}
\end{equation}
with upper index for the Dirac cone index.
Note that the inverse sign comes from the anomaly cancellation between the paired Dirac cones.
We arrive at 
\begin{equation}
    N = \mathrm{sgn}(\sigma^{\text{MD},1}_H) + \mathrm{sgn}(\sigma^{\text{MD},2}_H) = 0 = \sigma^{\text{MD},1}_H + \sigma^{\text{MD},2}_H.
\end{equation}
After the back deformation, we have invariant Chern number, while the anomalous Hall conductance is generally not protected to be zero, as 
\begin{equation}
    \begin{cases}
        N = \mathrm{sgn}(\sigma^{1}_H) + \mathrm{sgn}(\sigma^{2}_H) = 0\\
        \sigma_H = \sigma^{1}_H + \sigma^{2}_H \neq 0
    \end{cases}.
\end{equation}

This guideline can be generalized to multi-Fermi loop.
By labelling Fermi loop in $\{l_i\}_{i = 1}^n$,
the above procedure tells and generalizes the results to that 
\begin{equation}
    N = \sum_i \phi_i^{\text{MD}}/\pi = \sum_i 2 \sigma_H^{\text{MD},i},
\end{equation}
where $\phi_i^{\text{MD}}$ is 1D Berry phase at $i$-th Fermi loop defined using idealized massless Dirac cone, or equivalently the AHC $\sigma_H^{\text{MD},i}$ by Stokes theorem.
Note that here $\phi_i^{\text{MD}} = n \pi$ protected by the emergent time-reversal or parity symmetry at Fermi surface in idealized band structure\cite{fu2022half,Zou2023half,Bai2024Dirac,fu2024invariants,Shen2024Coshare},
and it is robust against gauge transformation, as it is defined on the boundary of 2D band-energy surface, and no large gauge transformation is allowed.
This formula is valid for both continuum and lattice models, and when the two models are consistent,
the formula degrades to the one appeared in the main text
\begin{equation}
    N = \sum_i |n_i| \mathrm{sgn}(\sigma_H^i),
\end{equation}
with $n_i$ the 1D winding number of the $i$-th Fermi loop defined using idealized massless Dirac cone, with its contribution fixed by the sign of the AHC $\sigma_H^i$.
Basically, a low-energy $k_-^{n_i} \sigma_+ + k_+^{n_i}\sigma_-$ gives winding number $n_i$.
Also note that when extracting the relevant bands by decoupling, we can not change the Chern number carried by the rest of the bands.
Meanwhile, there are cases with inseparable Fermi loops where they are assigned to the same band or bands with intersections.
For the former, gauge consistency of occupied states need to be checked (and at this time, Berry phase at each loop can be a better quantity),
while for the latter, one can perform idealization to isolate different bands without considering which parts of bands are connected, 
since we only care about the total Chern number.

\section{Experimental signature by temperature-dependent thermal Hall conductivity}\label{sec:THC}

\begin{figure}[htbp]
    \centering \includegraphics[width = 8.6cm]{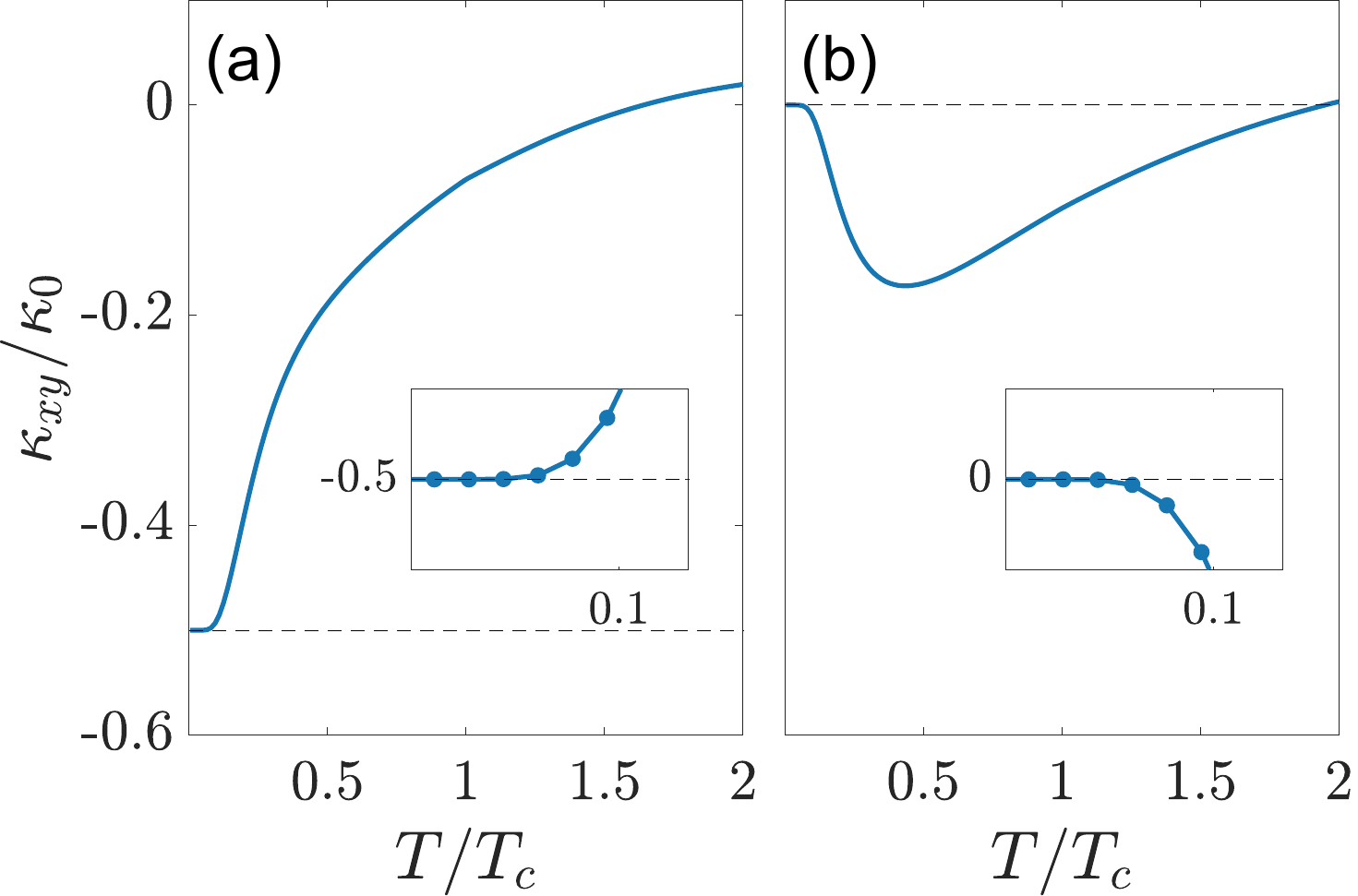} 
    \caption{Thermal Hall conductivity in unit of $\kappa_0 = \frac{\pi^2 k_B^2}{3h} T$ for (a) topological ($\mu = 0.06~\si{eV}$) and (b) trivial ($\mu = 0.12~\si{eV}$) superconducting phases, varying with temperature, in unit of superconducting gap.
    The inner panel shows the low-temperature behavior, highlighting the quantized value of thermal Hall conductivity.
    }
    \label{fig:TH_vary}
\end{figure}

Here, we propose to measure the temperature dependence of the thermal Hall conductivity\cite{Cooper1997Thermoelectric,Matsumoto2011Rotational,Qin2011energy} $\kappa_{xy}(T)$ as a feasible 
experimental signature of the CTSC phase.
As a heat analogue of electric Hall effect, the thermal Hall conductivity is expressed in the superconducting system\cite{Sumiyoshi2013Quantum} as
\begin{equation}
    \kappa_{xy} = -\frac{1}{2hT} \int \mathrm{d}\varepsilon~\varepsilon^2 N(\varepsilon) \frac{\mathrm{d} f(\varepsilon)}{\mathrm{d} \varepsilon},
\end{equation}
where the energy-dependent Berry curvature integral reads $N(\varepsilon) = 2\pi \sum_{\varepsilon_{n\bm{k}} < \varepsilon} \Omega_{n\bm{k}}/V$, with $\Omega_{n\bm{k}}$ the Berry curvature of $n$th BdG band and $V$ the system volume,
and $f$ is the Fermi-Dirac function.
In the low temperature limit, a Sommerfeld expansion leads to\cite{Read2000Paired} 
\begin{equation}
    \kappa_{xy}(T) = - \frac{\pi^2 k_B^2}{6h} N(0) T + O(T^3),
\end{equation}
with $k_B$, $h$ the Boltzmann constant and Planck constant, respectively.
Note that $N(0)$ is just the superconducting Chern number $N$, and leads to a quantized thermal Hall conductivity equals to $N/2$ in unit of $\kappa_0 = \frac{\pi^2 k_B^2}{3h} T$ when temperature is low enough.
On the other hand, as temperature increases, not only superconducting gap melts, but 
higher states contribute to the thermal Hall conductivity, making it generally deviates from the quantized value.
We implement this idea numerically within our system, with semi-empirical interpolation formula for superconducting gap\cite{Tachiki1991Superconducting}
\begin{equation}
    \Delta(T) \approx \Delta(0)\Theta(T_c-T)\tanh(1.74\sqrt{T_c/T - 1}),
\end{equation}
where $T_c \approx \Delta(0)/1.76k_B$ is the superconducting critical temperature, $\Delta(0) = \Delta$ is the zero-temperature gap, and $\Theta$ is the Heaviside step function.
As presented in Fig.~\ref{fig:TH_vary}, in a series of measurements of thermal Hall conductivity over our system with decreasing temperature,
we see a crossover from a non-quantized value at relatively high temperature to a quantized value at low temperature,
with the quantization regime generally bounded by the superconducting gap strength.
Especially, we have $\kappa_{xy}(T\rightarrow 0)/\kappa_0 \rightarrow N/2$ in the low temperature limit, making
it a clear signal for the presence or absence of the CTSC phase, as illustrated in Fig.~\ref{fig:TH_vary}(a) and (b), respectively.

\section{Discussion and summary}

Before concluding, we discuss the feasibility and broader implications
of the proposed system. In the normal bulk-like stacking case (left
side of Fig. 2), where the crystal orientations of the septuple layers
are identical, the system either exhibits inherent $\mathcal{P}\mathcal{T}$
symmetry for even septuple layers, or bears $\mathcal{M}_{z}$ symmetry
with vanishingly small finite-size effect, the system will thus always
carry vanishing polarization and degenerate bands, and always has
double Fermi loop. This is the reason why no evidence of CTSC was
measured\cite{Yuan2024Coexistence}. In contrast, the polar stacking
with a 180$^{\circ}$ rotated top layer and interlayer sliding (right
side of Fig. 2) breaks the $\mathcal{P}\mathcal{T}$ symmetry, leading
to ferroelectricity and band splitting. The sliding occurs spontaneously,
resulting in two stable configurations related by $\mathcal{M}_{z}\mathcal{T}$
symmetry with opposite polarizations, which can be tuned by an external
electric field or optical excitation\cite{Yang2024Light}.

The residual chirality guideline can also be generalized to multi-band
systems by the summation over all the Fermi loops: $N=\sum_{i}|n_i|\mathrm{sgn}(\sigma_{H}^{i})$, 
with $n_i$ the 1D Fermi loop winding number of $i$th Fermi loop.
In this context, we highlight several candidate normal-state systems
for realizing the CTSC phase. These include half-quantum Hall system\cite{Muzaffar2025Switchable}, non-collinear antiferromagnets
\cite{Chen2014Anomalous,Nakatsuji2015anomalous}, as well as collinear
antiferromagnets with band splittings induced by mechanisms such as
chiral crystal fields \cite{Smejkal2020AHEAFM,Takagi2024Hall} or
charge-density-wave order \cite{Zyuzin2024Anomalous}, among others.

In summary, we have demonstrated that the multiferroic polar-stacking
bilayer MnBi$_{2}$Te$_{4}$/Fe(Se,Te) heterostructure provides a
robust and experimentally feasible platform for realizing chirality-controllable
topological superconductivity.

\begin{acknowledgments}
K.-Z. Bai acknowledges helpful discussions with S. B. Zhang. This
work was supported by the Quantum Science Center of Guangdong-Hong
Kong-Macao Greater Bay Area (Grant No. GDZX2301005) and the Research
Grants Council, University Grants Committee, Hong Kong (Grants No.
C7012-21G and No. 17301823).
\end{acknowledgments}


\begin{thebibliography}{99}
\bibitem{Kitaev2001Unpaired} A. Y. Kitaev, Unpaired Majorana fermions in quantum wires, Physics-Uspekhi \textbf{44}, 131 (2001).
\bibitem{Fu2008Superconducting} L. Fu and C. L. Kane, Superconducting proximity effect and Majorana fermions at the surface of a topological insulator, Phys. Rev. Lett. \textbf{100}, 096407 (2008).
\bibitem{Lutchyn2010Majorana} R. M. Lutchyn, J. D. Sau, and S. Das Sarma, Majorana fermions and a topological phase transition in semiconductor-superconductor heterostructures, Phys. Rev. Lett. \textbf{105}, 077001 (2010).
\bibitem{Qi2011Topological} X.-L. Qi and S.-C. Zhang, Topological insulators and superconductors, Rev. Mod. Phys. \textbf{83}, 1057 (2011).
\bibitem{shen2012topological} S.-Q. Shen, Topological insulators, Vol. 174 (Springer, 2012).
\bibitem{Sato2017TSC} M. Sato and Y. Ando, Topological superconductors: a review, Reports on Progress in Physics \textbf{80}, 076501 (2017).
\bibitem{Read2000Paired} N. Read and D. Green, Paired states of fermions in two dimensions with breaking of parity and time-reversal symmetries and the fractional quantum Hall effect, Phys. Rev. B \textbf{61}, 10267 (2000).
\bibitem{Wilczek2009Majorana} F. Wilczek, Majorana returns, Nature Physics \textbf{5}, 614 (2009).
\bibitem{Qi2010Chiral} X.-L. Qi, T. L. Hughes, and S.-C. Zhang, Chiral topological superconductor from the quantum Hall state, Phys. Rev. B \textbf{82}, 184516 (2010).
\bibitem{Wang2015Chiral} J. Wang, Q. Zhou, B. Lian, and S.-C. Zhang, Chiral topological superconductor and half-integer conductance plateau from quantum anomalous Hall plateau transition, Phys. Rev. B \textbf{92}, 064520 (2015).
\bibitem{Fu2023Anomalous} B. Fu and S.-Q. Shen, Anomalous coherence length of Majorana zero modes at vortices in superconducting topological insulators, Phys. Rev. B \textbf{107}, 184517 (2023).
\bibitem{Bravyi2002Computation} S. B. Bravyi and A. Y. Kitaev, Fermionic quantum computation, Annals of Physics \textbf{298}, 210 (2002).
\bibitem{Kitaev2003TQC} A. Kitaev, Fault-tolerant quantum computation by anyons, Annals of Physics \textbf{303}, 2 (2003).
\bibitem{Nayak2008NonAbelian} C. Nayak, S. H. Simon, A. Stern, M. Freedman, and S. Das Sarma, Non-Abelian anyons and topological quantum computation, Rev. Mod. Phys. \textbf{80}, 1083 (2008).
\bibitem{Alicea2011TQIP} J. Alicea, Y. Oreg, G. Refael, F. von Oppen, and M. P. A. Fisher, Non-Abelian statistics and topological quantum information processing in 1D wire networks, Nature Physics \textbf{7}, 412 (2011).
\bibitem{Lian2018TQC} B. Lian, X.-Q. Sun, A. Vaezi, X.-L. Qi, and S.-C. Zhang, Topological quantum computation based on chiral Majorana fermions, Proceedings of the National Academy of Sciences \textbf{115}, 10938 (2018).
\bibitem{Fu20073DTI} L. Fu, C. L. Kane, and E. J. Mele, Topological insulators in three dimensions, Phys. Rev. Lett. \textbf{98}, 106803 (2007).
\bibitem{xia2009observation} Y. Xia, D. Qian, D. Hsieh, L. Wray, A. Pal, H. Lin, A. Bansil, D. Grauer, Y. S. Hor, R. J. Cava, and M. Z. Hasan, Observation of a large-gap topological-insulator class with a single Dirac cone on the surface, Nat. Phys. \textbf{5}, 398 (2009).
\bibitem{zhang2009topological} H. Zhang, C.-X. Liu, X.-L. Qi, X. Dai, Z. Fang, and S.-C. Zhang, Topological insulators in Bi2Se3, Bi2Te3 and Sb2Te3 with a single Dirac cone on the surface, Nat. Phys. \textbf{5}, 438 (2009).
\bibitem{chen2010massive} Y. L. Chen, J.-H. Chu, J. G. Analytis, Z. K. Liu, K. Igarashi, H.-H. Kuo, X. L. Qi, S. K. Mo, R. G. Moore, D. H. Lu, M. Hashimoto, T. Sasagawa, S. C. Zhang, I. R. Fisher, Z. Hussain, and Z. X. Shen, Massive Dirac fermion on the surface of a magnetically doped topological insulator, Science \textbf{329}, 659 (2010).
\bibitem{Haldane1988QAHE} F. D. M. Haldane, Model for a quantum Hall effect without Landau levels: Condensed-matter realization of the "parity anomaly", Phys. Rev. Lett. \textbf{61}, 2015 (1988).
\bibitem{Yu2010QAHE} R. Yu, W. Zhang, H.-J. Zhang, S.-C. Zhang, X. Dai, and Z. Fang, Quantized anomalous Hall effect in magnetic topological insulators, Science \textbf{329}, 61 (2010).
\bibitem{Chang2013QAHE} C.-Z. Chang, J. Zhang, X. Feng, J. Shen, Z. Zhang, M. Guo, K. Li, Y. Ou, P. Wei, L.-L. Wang, Z.-Q. Ji, Y. Feng, S. Ji, X. Chen, J. Jia, X. Dai, Z. Fang, S.-C. Zhang, K. He, Y. Wang, L. Lu, X.-C. Ma, and Q.-K. Xue, Experimental observation of the quantum anomalous Hall effect in a magnetic topological insulator, Science \textbf{340}, 167 (2013).
\bibitem{Stern2021Interfacial} M. V. Stern, Y. Waschitz, W. Cao, I. Nevo, K. Watanabe, T. Taniguchi, E. Sela, M. Urbakh, O. Hod, and M. B. Shalom, Interfacial ferroelectricity by van der Waals sliding, Science \textbf{372}, 1462 (2021).
\bibitem{Tsymbal2021Ferroelectric} K. Yasuda, X. Wang, K. Watanabe, T. Taniguchi, and P. Jarillo-Herrero, Stacking-engineered ferroelectricity in bilayer boron nitride, Science \textbf{372}, 1458 (2021).
\bibitem{Wang2022rhombohedral} X. Wang, K. Yasuda, Y. Zhang, S. Liu, K. Watanabe, T. Taniguchi, J. Hone, L. Fu, and P. Jarillo-Herrero, Interfacial ferroelectricity in rhombohedral-stacked bilayer transition metal dichalcogenides, Nature Nanotechnology \textbf{17}, 367 (2022).
\bibitem{Weston2022FEtwisted} A. Weston, E. G. Castanon, V. Enaldiev, F. Ferreira, S. Bhattacharjee, S. Xu, H. Corte-León, Z. Wu, N. Clark, A. Summerfield, T. Hashimoto, Y. Gao, W. Wang, M. Hamer, H. Read, L. Fumagalli, A. V. Kretinin, S. J. Haigh, O. Kazakova, A. K. Geim, V. I. Fal’ko, and R. Gorbachev, Interfacial ferroelectricity in marginally twisted 2D semiconductors, Nature Nanotechnology \textbf{17}, 390 (2022).
\bibitem{Gou2023ferroelectricity} J. Gou, H. Bai, X. Zhang, Y. L. Huang, S. Duan, A. Ariando, S. A. Yang, L. Chen, Y. Lu, and A. T. S. Wee, Two-dimensional ferroelectricity in a single-element bismuth monolayer, Nature \textbf{617}, 67 (2023).
\bibitem{Cao2023Switchable} T. Cao, D.-F. Shao, K. Huang, G. Gurung, and E. Y. Tsymbal, Switchable anomalous Hall effects in polar-stacked 2D antiferromagnet MnBi2Te4, Nano Letters \textbf{23}, 3781 (2023).
\bibitem{Ren2022MBT} Y. Ren, S. Ke, W.-K. Lou, and K. Chang, Quantum phase transitions driven by sliding in bilayer MnBi2Te4, Phys. Rev. B \textbf{106}, 235302 (2022).
\bibitem{Luo2023MBT} W. Luo, M.-H. Du, F. A. Reboredo, and M. Yoon, Non-volatile electric control of magnetic and topological properties of MnBi2Te4 thin films, 2D Materials \textbf{10}, 035008 (2023).
\bibitem{Muzaffar2025Switchable} M. U. Muzaffar, K.-Z. Bai, W. Qin, G. Cao, B. Fu, P. Cui, S.-Q. Shen, and Z. Zhang, Ferroelectrically switchable half-quantized Hall effect, Nano Letters (2025).
\bibitem{Scott2007Ferroelectrics} J. F. Scott, Applications of modern ferroelectrics, Science \textbf{315}, 954 (2007).
\bibitem{Young2012Shift} S. M. Young and A. M. Rappe, First principles calculation of the shift current photovoltaic effect in ferroelectrics, Phys. Rev. Lett. \textbf{109}, 116601 (2002).
\bibitem{Dagdeviren2014response} C. Dagdeviren, Y. Su, P. Joe, R. Yona, Y. Liu, Y.-S. Kim, Y. Huang, A. R. Damadoran, J. Xia, L. W. Martin, Y. Huang, and J. A. Rogers, Conformable amplified lead zirconate titanate sensors with enhanced piezoelectric response for cutaneous pressure monitoring, Nature Communications \textbf{5}, 4496 (2014).
\bibitem{Eerenstein2006Multiferroic} W. Eerenstein, N. D. Mathur, and J. F. Scott, Multiferroic and magnetoelectric materials, Nature \textbf{442}, 759 (2006).
\bibitem{Dong2015Multiferroic} S. Dong, J.-M. Liu, S.-W. Cheong, and Z. Ren, Multiferroic materials and magnetoelectric physics: symmetry, entanglement, excitation, and topology, Advances in Physics \textbf{64}, 519 (2015).
\bibitem{He2014interface} Q. L. He, H. Liu, M. He, Y. H. Lai, H. He, G. Wang, K. T. Law, R. Lortz, J. Wang, and I. K. Sou, Two-dimensional superconductivity at the interface of a Bi2Te3/FeTe heterostructure, Nature Communications \textbf{5}, 4247 (2014).
\bibitem{Owada2019Electronic} K. Owada, K. Nakayama, R. Tsubono, K. Shigekawa, K. Sugawara, T. Takahashi, and T. Sato, Electronic structure of a Bi2Te3/FeTe heterostructure: Implications for unconventional superconductivity, Phys. Rev. B \textbf{100}, 064518 (2019).
\bibitem{Qin2020Superconductivity} H. Qin, B. Guo, L. Wang, M. Zhang, B. Xu, K. Shi, T. Pan, L. Zhou, J. Chen, Y. Qiu, B. Xi, I. K. Sou, D. Yu, W.-Q. Chen, H. He, F. Ye, J.-W. Mei, and G. Wang, Superconductivity in single-quintuple-layer Bi2Te3 grown on epitaxial FeTe, Nano Letters \textbf{20}, 3160 (2020).
\bibitem{Ding2022Observation} S. Ding, C. Chen, Z. Cao, D. Wang, Y. Pan, R. Tao, D. Zhao, Y. Hu, T. Jiang, Y. Yan, Z. Shi, X. Wan, D. Feng, and T. Zhang, Observation of robust zero-energy state and enhanced superconducting gap in a trilayer heterostructure of MnTe/Bi2Te3/Fe(Te, Se), Science Advances \textbf{8}, eabq4578 (2022).
\bibitem{Yuan2024Coexistence} W. Yuan, Z.-J. Yan, H. Yi, Z. Wang, S. Paolini, Y.-F. Zhao, L. Zhou, A. G. Wang, K. Wang, T. Prokscha, Z. Salman, A. Suter, P. P. Balakrishnan, A. J. Grutter, L. E. Winter, J. Singleton, M. H. W. Chan, and C.-Z. Chang, Coexistence of superconductivity and antiferromagnetism in topological magnet MnBi2Te4 films, Nano Letters \textbf{24}, 7962 (2024).
\bibitem{Xiao2010Berry} D. Xiao, M.-C. Chang, and Q. Niu, Berry phase effects on electronic properties, Rev. Mod. Phys. \textbf{82}, 1959 (2010).
\bibitem{Nagaosa2010AHE} N. Nagaosa, J. Sinova, S. Onoda, A. H. MacDonald, and N. P. Ong, Anomalous Hall effect, Rev. Mod. Phys. \textbf{82}, 1539 (2010).
\bibitem{haldane2004berry} F. D. M. Haldane, Berry curvature on the Fermi surface: Anomalous Hall effect as a topological Fermi-liquid property, Phys. Rev. Lett. \textbf{93}, 206602 (2004).
\bibitem{winkler2003spin} R. Winkler, S. Papadakis, E. De Poortere, and M. Shayegan, Spin-orbit coupling in two-dimensional electron and hole systems, Vol. 41 (Springer, 2003).
\bibitem{Liu2010Model} C.-X. Liu, X.-L. Qi, H. Zhang, X. Dai, Z. Fang, and S.-C. Zhang, Model Hamiltonian for topological insulators, Phys. Rev. B \textbf{82}, 045122 (2010).
\bibitem{Yang2014Classification} B.-J. Yang and N. Nagaosa, Classification of stable three-dimensional Dirac semimetals with nontrivial topology, Nature Communications \textbf{5}, 4898 (2014).
\bibitem{Acosta2018Tightbinding} C. M. Acosta, M. P. Lima, A. J. R. da Silva, A. Fazzio, and C. H. Lewenkopf, Tight-binding model for the band dispersion in rhombohedral topological insulators over the whole Brillouin zone, Phys. Rev. B \textbf{98}, 035106 (2018).
\bibitem{fu2024invariants} B. Fu and S.-Q. Shen, $\mathbb{Z}/2$ topological invariants and the half quantized Hall effect, Commun Phys \textbf{8}, 2 (2025).

\bibitem{Fu2009Hexagonal} L. Fu, Hexagonal warping effects in the surface states of the topological insulator Bi2Te3, Phys. Rev. Lett. \textbf{103}, 266801 (2009).
\bibitem{Note-on-SM} See Supplemental Material at [URL to be added by publisher] for details of construction of the eﬀective models and discussion over superconducting single particle properties, which includes Refs. \cite{Fu2009Hexagonal,Liu2010Model,Bai2024Dirac,Bai2023Metallic,Shen2017Topological,Alicea2010Majorana}.
\bibitem{Bai2023Metallic} K.-Z. Bai, B. Fu, Z. Zhang, and S.-Q. Shen, Metallic quantized anomalous Hall effect without chiral edge states, Phys. Rev. B \textbf{108}, L241407 (2023).
\bibitem{Bai2024Dirac} K.-Z. Bai, B. Fu, and S.-Q. Shen, Dirac fermions and topological phases in magnetic topological insulator films, SciPost Phys. \textbf{17}, 146 (2024).
\bibitem{mahan2000many} G. D. Mahan, Nonzero temperatures, in Many-Particle Physics (Springer US, Boston, MA, 2000) pp. 109–185.
\bibitem{Kitaev2009Periodic} A. Kitaev, Periodic table for topological insulators and superconductors, AIP Conference Proceedings \textbf{1134}, 22 (2009).
\bibitem{Ryu2010tenfold} S. Ryu, A. P. Schnyder, A. Furusaki, and A. W. W. Ludwig, Topological insulators and superconductors: tenfold way and dimensional hierarchy, New Journal of Physics \textbf{12}, 065010 (2010).
\bibitem{Chiu2016Classification} C.-K. Chiu, J. C. Y. Teo, A. P. Schnyder, and S. Ryu, Classification of topological quantum matter with symmetries, Rev. Mod. Phys. \textbf{88}, 035005 (2016).
\bibitem{Takahashi2002Hall} S. Takahashi and S. Maekawa, Hall effect induced by a spin-polarized current in superconductors, Phys. Rev. Lett. \textbf{88}, 116601 (2002).
\bibitem{Sacramento2012Anomalous} P. D. Sacramento, M. A. N. Araújo, V. R. Vieira, V. K. Dugaev, and J. Barnaś, Anomalous Hall effect in superconductors with spin-orbit interaction, Phys. Rev. B \textbf{85}, 014518 (2012).
\bibitem{POjanen2013Anomalous} T. Ojanen and T. Kitagawa, Anomalous electromagnetic response of superconducting Rashba systems in trivial and topological phases, Phys. Rev. B \textbf{87}, 014512 (2013).
\bibitem{Bednik2016Anomalous} G. Bednik, A. A. Zyuzin, and A. A. Burkov, Anomalous Hall effect in Weyl superconductors, New Journal of Physics \textbf{18}, 085002 (2016).
\bibitem{Sacramento2014Hall} P. D. Sacramento, M. A. N. Araújo, and E. V. Castro, Hall conductivity as bulk signature of topological transitions in superconductors, Europhysics Letters \textbf{105}, 37011 (2014).
\bibitem{Zou2023half} J.-Y. Zou, R. Chen, B. Fu, H.-W. Wang, Z.-A. Hu, and S.-Q. Shen, Half-quantized Hall effect at the parity-invariant Fermi surface, Phys. Rev. B \textbf{107}, 125153 (2023).
\bibitem{Shen2024Coshare} S. Q. Shen, CoShare Science 2, 1 (2024).
\bibitem{fu2022half} B. Fu, J.-Y. Zou, Z.-A. Hu, H.-W. Wang, and S.-Q. Shen, Quantum anomalous semimetals, npj Quantum Mater. \textbf{7}, 94 (2022).
\bibitem{Cooper1997Thermoelectric} N. R. Cooper, B. I. Halperin, and I. M. Ruzin, Thermoelectric response of an interacting two-dimensional electron gas in a quantizing magnetic field, Phys. Rev. B \textbf{55}, 2344 (1997).
\bibitem{Matsumoto2011Rotational} R. Matsumoto and S. Murakami, Rotational motion of magnons and the thermal Hall effect, Phys. Rev. B \textbf{84}, 184406 (2011).
\bibitem{Qin2011energy} T. Qin, Q. Niu, and J. Shi, Energy magnetization and the thermal Hall effect, Phys. Rev. Lett. \textbf{107}, 236601 (2011).
\bibitem{Sumiyoshi2013Quantum} H. Sumiyoshi and S. Fujimoto, Quantum thermal Hall effect in a time-reversal-symmetry-broken topological superconductor in two dimensions: Approach from bulk calculations, Journal of the Physical Society of Japan \textbf{82}, 023602 (2013).
\bibitem{Tachiki1991Superconducting} M. Tachiki, T. Koyama, and S. Takahashi, Superconducting properties in layered cuprate oxides, in Advances in Superconductivity III (Springer, 1991) pp. 45–51.
\bibitem{Yang2024Light} Q. Yang and S. Meng, Light-induced complete reversal of ferroelectric polarization in sliding ferroelectrics, Phys. Rev. Lett. \textbf{133}, 136902 (2024).
\bibitem{Chen2014Anomalous} H. Chen, Q. Niu, and A. H. MacDonald, Anomalous Hall effect arising from noncollinear antiferromagnetism, Phys. Rev. Lett. \textbf{112}, 017205 (2014).
\bibitem{Nakatsuji2015anomalous} S. Nakatsuji, N. Kiyohara, and T. Higo, Large anomalous Hall effect in a non-collinear antiferromagnet at room temperature, Nature \textbf{527}, 212 (2015).
\bibitem{Smejkal2020AHEAFM} L. Šmejkal, R. González-Hernández, T. Jungwirth, and J. Sinova, Crystal time-reversal symmetry breaking and spontaneous Hall effect in collinear antiferromagnets, Science Advances \textbf{6}, eaaz8809 (2020).
\bibitem{Takagi2024Hall} R. Takagi, R. Hirakida, Y. Settai, R. Oiwa, H. Takagi, A. Kitaori, K. Yamauchi, H. Inoue, J.-i. Yamaura, D. Nishio-Hamane, S. Itoh, S. Aji, H. Saito, T. Nakajima, T. Nomoto, R. Arita, and S. Seki, Spontaneous Hall effect induced by collinear antiferromagnetic order at room temperature, Nature Materials (2024).
\bibitem{Zyuzin2024Anomalous} V. A. Zyuzin, Anomalous Hall effect in metallic collinear antiferromagnets with charge density wave order, Phys. Rev. B \textbf{110}, 174426 (2024).

\bibitem{Shen2017Topological} S.-Q. Shen, Topological invariants, in Topological Insulators: Dirac Equation in Condensed Matters (Springer, 2017) pp. 51-79.
\bibitem{Alicea2010Majorana} J. Alicea, Majorana fermions in a tunable semiconductor device, Phys. Rev. B \textbf{81}, 125318 (2010).
\end{thebibliography}

\end{document}